   \newcommand{\h}[2]{\mbox{\footnotesize$\frac{#1}{#2}$\normalsize}}
   \newcommand{\be}[0]{\begin{equation}}
   \newcommand{\ee}[0]{\end{equation}}
   \newcommand{\ba}[0]{\begin{eqnarray}}
   \newcommand{\ea}[0]{\end{eqnarray}}

        \newcommand{\BC}{{\hbox{\rm\vphantom{X}%
                       \hskip 0.25em%
                       \vrule width 0.7pt%
                       \hskip -0.35em C}}}
\documentstyle[12pt]{article}
\begin{document}
\Large
\hfill\vbox{\hbox{DTP/94/58}
            \hbox{September 1994}}
\nopagebreak

\vspace{0.75cm}
\begin{center}
\LARGE
{\bf Renormalon Singularities of the QCD Vacuum Polarization Function
to Leading Order in $1/N_{f}$}
\vspace{0.6cm}
\Large

C.N.Lovett-Turner and C.J.Maxwell

\vspace{0.4cm}
\large
\begin{em}
Centre for Particle Theory, University of Durham\\
South Road, Durham, DH1 3LE, England
\end{em}

\vspace{1.7cm}

\end{center}
\normalsize
\vspace{0.45cm}

We explicitly determine the residues and orders of all the
ultra-violet (UV) and infra-red (IR) renormalon poles in the
Borel plane
for the QCD vacuum polarization function (Adler D-function), to
leading order in an expansion in the number of quark flavours,
$N_{f}$. The singularity structure is precisely as anticipated on
general grounds. In particular, the leading IR renormalon is
absent, in agreement
with operator product expansion ideas. There is a curious and
unexplained symmetry between the third and higher UV and IR
renormalon residues. We are able to sum up separately UV and IR
contributions to obtain closed form results involving
$\zeta$-functions. We argue that the leading UV renormalon should
have a more complicated structure than conventionally assumed.
The disappearance of IR renormalons
in flavour-saturated SU($N$) QCD is shown to
occur for $N=3,6$ or 9.
\newpage
\section{Introduction}

The topic of the large-order behaviour of the expansion coefficients
in perturbative field theory calculations continues to be vigorously
pursued in the literature [1--9]. Notwithstanding the importance
of
establishing to what extent physical observables can be formally
reconstructed from their divergent perturbative expansions, there is
the more pragmatic motivation of the need to assess the reliability
of the growing number of higher-order perturbative calculations in
QCD and QED.

In this paper we wish to make use of recent progress \cite{f,h} in
exact
all-orders QED calculations to leading order in the $1/N_{f}$
expansion, with $N_{f}$ the number of fermions, to explicitly
determine the singularity structure of the perturbative QCD vacuum
polarization function in the Borel plane. On very general grounds
one anticipates branch point singularities evenly spaced along the
positive and negative real axis in the Borel variable \cite{b}.
Those on
the positive axis, referred to as infra-red (IR) renormalons, are
supposedly correlated with the absence from the formal perturbation
series of infra-red non-perturbative effects, vacuum condensates,
present in the operator product expansion (OPE). They are
responsible for fixed-sign factorial growth of the series
coefficients and represent a genuine ambiguity in reconstructing
the physical observable from the formal perturbation series. Those
on the negative axis, so-called ultra-violet (UV) renormalons,
correspond to alternating-sign factorial growth of the series
coefficients and do not prevent the reconstruction of the
observable by Borel summation.

Whilst the above singularity structure is well-motivated
theoretically, there have been various problematic issues. In
particular the connection with the OPE suggests that the leading IR
renormalon should be absent for the case of the QCD vacuum
polarization function since there is no relevant operator of
dimension two; the first contribution being the gluon condensate of
dimension four. This conclusion has been questioned on various
grounds by several authors \cite{j,k}.

The leading asymptotic growth of the perturbative coefficients will
be determined by the Borel plane singularity nearest the origin; for
the case of the QCD vacuum polarization function this is the first
UV renormalon. We point out that the conventionally expected
structure of this singularity, with a single branch point exponent,
would enable one to obtain the asymptotic growth of the coefficients
to all orders in the $1/N_{f}$ expansion given an exact
large-$N_{f}$ result. We suggest that this is unlikely and indicate
a more complicated structure for the first UV renormalon in
accordance with recent results of Vainshtein and Zakharov obtained
using their ``UV renormalon calculus'' \cite{g}.

We shall show that the actual
singularity structure of the QCD vacuum polarization function is
precisely as expected; in particular the leading IR renormalon
singularity is indeed absent. This has also been noted for the
singularities in the QED vacuum polarization function in reference
\cite{f}.
We further demonstrate that there is an unexpected symmetry between
the third and higher UV and IR singularities. We are able to sum up
the UV and IR contributions separately to obtain a closed form
result involving $\zeta$-functions.

We finally
show that in SU(3) QCD, with $N_{f}$=15 or 16, the IR renormalon
singularities are absent \cite{l,m}; and that they first vanish when
the
instanton/anti-instanton singularity becomes leading. The
requirement that this happens for an SU($N$) theory uniquely selects
$N=3$.

The organisation of the paper is as follows. In section 2 we shall
introduce the Adler D-function and define its $1/N_{f}$ expansion. We
shall also discuss the exact large-$N_{f}$ QCD result for its
coefficients \cite{i}. In section 3 we define the Borel plane and
review
the anticipated singularity structure.
We discuss reasons for which one may expect a more complicated
singularity structure for the first UV renormalon and
make contact with the results of reference \cite{g}. In section 4 we
shall
explicitly determine the Borel plane singularity structure for the
exact large-$N_{f}$ result for D; and show that the anticipated UV
and IR renormalon poles are present. In section 5 we discuss the
vanishing of the IR renormalons for certain values of $N_{f}$ in
SU($N$) QCD. Section 6 contains our conclusions.
\section{The Adler D-function and the $1/N_{f}$ Expansion}

We shall be interested in the SU($N$) QCD vacuum polarization
function with $N_{f}$ flavours of massless quarks,

\be
\Pi(-q^{2})(q_{\mu}q_{\nu}-g_{\mu\nu}q^{2})=16\pi^{2}i
\int\mbox{d}^{4}x\,e^{i{\bf q.x}}\langle 0|T\{J_{\mu}(x)
J_{\nu}(0)\}|0\rangle
\;.
\ee

To avoid an unspecified constant we shall actually focus on the
related Adler D-function,
\be
D(Q^{2})=-\frac{3}{4}Q^{2}\frac{\mbox{d}}{\mbox{d}Q^{2}}\Pi(Q^{2})
\ee
where $Q^{2}=-q^{2}$ is the spacelike Euclidean squared momentum
transfer. This quantity is related to the experimentally-relevant
R-ratio in e$^{+}$e$^{-}$ annihilation,
\be
R=\frac{\sigma(\mbox{e}^{+}\mbox{e}^{-}\rightarrow \mbox{hadrons})}
{\sigma(\mbox{e}^{+}\mbox{e}^{-}\rightarrow \mu^{+}\mu^{-})}
\ee
where, taking $s$ to be the physical timelike Minkowski squared
momentum transfer, R and D are related by the dispersion relation
\be
R(s)=\frac{1}{2\pi i}\int_{-s-i\epsilon}^{-s+i\epsilon}\mbox{d}
Q^{2}
\frac{D(Q^{2})}{Q^{2}}\;.
\ee
In QCD perturbation theory we have
\be
D(Q^{2})=d(R)\sum_{f}Q_{f}^{2}\left(1+\frac{3}{4}C_{F}\tilde{D}\right)
+\left(\sum_{f}Q_{f}
\right)^{2}\tilde{\tilde{D}}
\ee
where $Q_{f}$ denotes the electric charge of the quarks and the
summation is over the flavours accessible at a given energy. $d(R)$
is the dimension of the quark representation of the colour group
(here $d(R)=N$). We define the SU($N$) Casimirs $C_{A}=N$, $C_{F}=
(N^{2}-1)/2N$.

The correction to the parton model result has the perturbative
expansion
\be
\tilde{D}=a+d_{1}a^{2}+d_{2}a^{3}+\ldots+
d_{k}a^{k+1}+\ldots\;,
\ee
with `$a$' the renormalization group (RG) improved coupling
$\alpha_{s}
(\mu^{2})/\pi$. The $\tilde{\tilde{D}}$ contribution first enters at
O($a^{3}$) due to the existence of diagrams of the ``light-by-light''
type. Our interest here is in the asymptotic growth of the $d_{k}$
coefficients in large orders.

The RG-improved coupling $a(\mu^{2}$)
will evolve with renormalization scale $\mu^{2}$ according to the
beta-function equation
\be
\frac{\mbox{d}a}{\mbox{d}\ln\mu}=-ba^{2}(1+ca+c_{2}a^{2}+\ldots)
\ee
where $b$ and $c$ are universal with \cite{n}
\ba
b&=&\frac{(11C_{A}-2N_{f})}{6}\nonumber\\
c&=&\left[-\frac{7}{8}\frac{C_{A}^{2}}{b}-\frac{11}{8}\frac{C_{A}
C_{F}}{b}
+\frac{5}{4}C_{A}+\frac{3}{4}C_{F}\right]\;,
\ea
$c_{2}$ and higher coefficients are renormalization scheme (RS)
dependent. We shall usually consider the $\overline{\mbox{MS}}$
scheme with $\mu^{2}=Q^{2}$. For the R-ratio there is an analogous
expansion for the quantity $\tilde{R}$ with perturbative
coefficients $r_{k}$, defined as in equations (5) and (6). The
dispersion relation (4) means that the $r_{k}$ are directly related
to the $d_{k}$. For instance $r_{1}=d_{1}$ and
$r_{2}=d_{2}-\pi^{2}b^{2}/12$.
The $\pi^{2}$ terms arise due to analytic continuation.
Given knowledge of the asymptotic growth of the $d_{k}$ one can
obtain that of the $r_{k}$ using equation (4). We shall continue to
focus on the $d_{k}$ for the moment.

The coefficients $d_{k}$ will be polynomials of degree $k$ in
$N_{f}$:
\be
d_{k}=d_{k}^{[k]}N_{f}^{k}+d_{k}^{[k-1]}N_{f}^{k-1}
+\ldots+d_{k}^{[0]}\;,
\ee
where each term is a sum of multinomials in $C_{A}$, $C_{F}$ and
$N_{f}$ of
degree $k$ so that $d_{k}^{[r]}$ has the structure $C_{A}^{k-r-s}
C_{F}^{s}$ (note the prefactor of $C_{F}$ in equation (5)). The first
two coefficients $d_{1}$ and $d_{2}$ have been computed \cite{o}
and the
result using the $\overline{\mbox{MS}}$ scheme with $\mu^{2}=Q^{2}$,
expanded in $N_{f}$ as in equation (9), is
\ba
d_{1}&=&\left(-\frac{11}{12}+\frac{2}{3}\zeta_{3}\right)N_{f}
+C_{A}\left(\frac{41}{8}-\frac{11}{3}\zeta_{3}\right)
-\frac{1}{8}C_{F}\nonumber\\
d_{2}&=&\left(\frac{151}{162}-\frac{19}{27}\zeta_{3}\right)N_{f}^{2}
+C_{A}\left(-\frac{970}{81}+\frac{224}{27}\zeta_{3}+\frac{5}{9}
\zeta_{5}\right)N_{f}\nonumber\\
& &+C_{F}\left(-\frac{29}{96}+\frac{19}{6}\zeta_{3}-\frac{10}{3}
\zeta_{5}\right)N_{f}
+C_{A}^{2}\left(\frac{90445}{2592}-\frac{2737}{108}\zeta_{3}
-\frac{55}{18}\zeta_{5}\right)\nonumber\\
& &+C_{A}C_{F}\left(-\frac{127}{48}-\frac{143}{12}\zeta_{3}
+\frac{55}{3}\zeta_{5}\right)+C_{F}^{2}\left(-\frac{23}{32}\right)\;.
\ea
Here $\zeta_{p}$ denotes the Riemann zeta function,
\be
\zeta_{p}\equiv\sum_{l=1}^{\infty}l^{-p}\;.
\ee

$1/N_{f}$ expansions as in equation (9) have been widely used in the
past in the investigation of large-order behaviour and renormalons
\cite{f}.
As we shall emphasise, it is actually more useful for these purposes
to consider an expansion in powers of $b$. We can write
\be
d_{k}=d_{k}^{(k)}b^{k}+d_{k}^{(k-1)}b^{k-1}+\ldots+d_{k}^{(0)}\;.
\ee
The leading coefficient in the $1/N_{f}$ expansion is exactly related
to that in the ``$1/b$ expansion'' with
\be
d_{k}^{[k]}=(-1/3)^{k}d_{k}^{(k)}\;.
\ee
As before, we can write out the known $d_{1}$ and $d_{2}$
coefficients
now expanded according to equation (12):
\ba
d_{1} &=& \left( \frac{11}{4}-2\zeta_{3}\right) b+\frac{C_{A}}{12}-
\frac{C_{F}}
{8}\nonumber\\
d_{2} &=& \left( \frac{151}{18}-\frac{19}{3}\zeta_{3}\right)b^{2}
+C_{A}\left(\frac{31}{6}-\frac{5}{3}\zeta_{3}-\frac{5}{3}
      \zeta_{5}
      \right) b\nonumber\\
      & & +C_{F}\left(\frac{29}{32}-\frac{19}{2}\zeta_{3}+10\zeta_{5}
      \right)b
      +C_{A}^{2}\left(-\frac{799}{288}-\zeta_{3}\right)\nonumber\\
      & & +C_{A}C_{F}\left(-\frac{827}{192}+\frac{11}{2}\zeta_{3}
      \right)+C_{F}^{2}\left(-\frac{23}{32}\right)\;.
\ea
We note in passing that the ``$1/b$ expansion'' has a somewhat more
compact structure than the $1/N_{f}$. In particular the $\zeta_{3}$
terms in $d_{1}$, which are present in all orders of the $1/N_{f}$
expansion, are now present only in the leading term in the
``$1/b$ expansion''; and the $\zeta_{5}$ terms present in all but the
leading coefficient in the $1/N_{f}$ expansion are now present only
in the $d_{2}^{(1)}$ coefficient in the ``$1/b$ expansion''.

Continuing progress in applying the $1/N_{f}$ expansion in QED
\cite{p} has
led Broadhurst to an elegant generating function for the leading
order (large-$N_{f}$) coefficients of the QED Gell-Mann--Low function
(MOM scheme beta-function) \cite{h}.
\be
\Psi_{n}^{[n]}=\frac{3^{2-n}}{2}\left(\frac{\mbox{d}}
{\mbox{d}x}\right)
^{n-2}P(x)
\biggr|_{x=1}
\ee
where
\be
P(x)=\frac{32}{3(1+x)}\sum_{k=2}^{\infty}\frac{(-1)^{k}k}{(k^{2}
-x^{2})^{2}}\;.
\ee
$\Psi_{n}^{[n]}$ can be explicitly evaluated in closed form \cite{h}
\ba
\frac{\Psi_{n}^{[n]}}{(n-2)!}&=&\frac{(n-1)}{(-3)^{n-1}}\biggr[
-2n+4-\frac{n+4}{2^{n}}\nonumber\\
& &+\frac{16}{n-1}\sum_{\h{n}{2}>s>0}s
(1-2^{-2s})(1-2^{2s-n})\zeta_{2s+1}\biggr]\;.
\ea
Using this result one can then obtain the leading-order
large-$N_{f}$ result for the QCD Adler D-function. In the $\overline
{\mbox{MS}}$ scheme with $\mu^{2}=Q^{2}$ one has \cite{i}
\be
d_{k}^{[k]}=2T_{f}^{k}k!\sum_{m=0}^{k}\frac{(-\h{5}{9})
^{m}}{m!}\frac{\Psi_{k+2-m}^{[k+2-m]}}{(k-m)!}\;,
\ee
where $T_{f}$ is a group theory factor; $T_{f}=1/2$ for the
standard fermion representation. The $(-5/9)^{m}$ factors
enter since one is converting from the MOM scheme Adler
function to that in the $\overline{\mbox{MS}}$ scheme. The results of
(18) and (17) are in agreement with the exactly known coefficients
$d_{1}^{[1]}$ and $d_{2}^{[2]}$ in equation (10).

Our aim is to make
use of the exact large-$N_{f}$ result of equation (18) to obtain as
much information as possible about the singularity structure of the
QCD D-function in the Borel plane; and hence about the large-order
behaviour of its perturbative coefficients. To this end we shall
begin by reviewing what can be inferred on very general grounds about
this structure; and then we shall compare the exact result with
these expectations.

\section{The Borel Plane and Renormalons}

Consider a general quantity $D$, calculated in perturbative field
theory with coupling $a$,
\be
D=(a+d_{1}a^{2}+d_{2}a^{3}+\ldots+d_{k}a^{k+1}+\ldots)
\ee
Using the integral representation of $k!$
\be
k!=\int_{0}^{\infty}dt\,e^{-t}t^{k}
\ee
we can formally write
\be
D=\sum_{m=0}^{\infty}a^{m+1}\frac{d_{m}}{m!}\int_{0}^{\infty}dt\,
e^{-t}t^{m}
\ee
with $d_{0}=1$. Then exchanging the order of summation and integration
\ba
D&=&\int_{0}^{\infty}\mbox{d}t\,ae^{-t}\sum_{m=0}^{\infty}
(ta)^{m}\frac
{d_{m}}{m!}\nonumber\\
&=&\int_{0}^{\infty}\mbox{d}z\,e^{-z/a}\sum_{m=0}^{\infty}\frac
{z^{m}d_{m}}{m!}\nonumber\\
&=&\int_{0}^{\infty}\mbox{d}z\,e^{-z/a}B[D](z)
\ea
where we have set $z=ta$. $B[D](z)$ is called the Borel transform of
$D(a)$ and is defined by
\be
B[D](z)\equiv\sum_{m=0}^{\infty}\frac{z^{m}d_{m}}{m!}\;.
\ee
The idea will be that even if the coefficients $d_{k}$ grow like
$k!$, as is expected in field theory, the Borel transform defined in
equation (23) may still be defined and hence the integral
representation of equation (22), the Borel sum, may exist.

As an example suppose that $d_{k}=(-1)^{k}k!$; then $B[D](z)=1-z+z^{2}
-z^{3}+\ldots=1/(1+z)$, where we have assumed analytic continuation
along the whole real line. More generally, if $d_{k}=(1/z_{i})^{k}
k^{\gamma}k!$ ($\gamma>0$), then $B[D](z)$ has a singularity
proportional to $(z-z_{i})^{-\gamma-1}$; so if $\gamma$ is a positive
integer we have a pole; and for non-integer $\gamma$ a branch point
in the $z$-plane (Borel plane) at $z=z_{i}$. If all the singularities
are located off the positive $z$-axis it may be possible, if certain
analyticity properties of $D$ are satisfied, to reconstruct $D(a)$
from its formal divergent series using the Borel sum of equation
(22). This, however, is not our interest here. We want to use the
$z$-plane singularities to encode the large-order behaviour of the
perturbative coefficients.

Returning to the specific example of the Adler D-function in QCD, let
us now ask on rather general grounds where the singularities $z_{i}$
could be located. In the large-$N_{f}$ limit we know that, if the
$d_{k}$ have factorial growth, asymptotically we must have
$d_{k}\approx N_{f}^{k}k!$( ). Similarly we can consider a large-$N$
limit where necessarily $d_{k}\approx N^{k}k!$( ). This follows
since the $k^{th}$ order Feynman diagrams for $d_{k}$
necessarily have factors which are multinomials in $N_{f}$, $C_{A}$
and $C_{F}$ of degree $k$. We therefore have singularities at
positions $z_{i}\sim 1/N_{f}$ in the large-$N_{f}$ limit and
$z_{i}\sim 1/N$ in the large-$N$ limit. If singularities are present
and visible in both limits then the simplest possibility is
$z_{i}\sim 1/(AN+BN_{f})$ involving some unspecified linear
combination of $N$ and $N_{f}$. These are the renormalon
singularities and in fact they lie at $z_{k}=2k/b$ where
$k=\pm 1,\pm 2,\pm 3,\ldots$. There are also singularities due to
instanton/anti-instanton solutions of the classical equations of
motion \cite{b,q,r}. These lie at $z_{k}=4k$, $k=1,2,3,\ldots$.
Since their positions are independent of $N$ and $N_{f}$, the above
arguments suggest that they are invisible in the large-$N$ and
large-$N_{f}$ limits. In fact they are invisible at all orders of the
$1/N$ and $1/N_{f}$ expansions, so we shall not learn about them
from the exact large-$N_{f}$ result.

We can motivate the $2/b$ spacing of the renormalons by considering
the connection between the IR renormalons and the absence from
the formal perturbation theory of non-perturbative vacuum
condensates present in the OPE. Performing a short distance OPE on
the time-ordered product of currents in equation (1) one has
\be
\tilde{D}(Q^{2})=C_{PT}(Q^{2}/\mu^{2},a(\mu^{2}))+
C_{GG}(Q^{2}/\mu^{2},a(\mu^{2})).\frac{\langle0|GG|0\rangle(\mu^{2})
}{Q^{4}}+O(Q^{-6}),
\ee
where $C_{PT}$ corresponds to the perturbation series of equation
(6). There are then higher terms consisting of perturbatively
calculable coefficient functions multiplied by non-perturbative
vacuum condensates of increasing mass dimension. The lowest such is
the gluon condensate with dimension 4 and a $Q^{-4}$ scaling
behaviour. There will also be dimension $6,8,\ldots$ condensates
corresponding to $Q^{-6},Q^{-8},\ldots$ behaviour. Crucially, there
is no relevant dimension 2 operator and hence no condensate with
$Q^{-2}$ behaviour. These condensates are supposedly associated with
the IR renormalon singularities at $z=4/b,6/b,\ldots$. The
connection follows from a consideration of the $Q^{2}$ behaviour of
the large-order perturbative coefficients $d_{k}$. Suppose that, for
$\mu^{2}=Q^{2}$, $d_{k}$ behaves asymptotically as
\be
d_{k}\approx A\left(\frac{1}{z_{i}}\right)^{k}k!(\mbox{ })\;,
\ee
where $A$ is a constant and the bracket denotes possible additional
powers or logarithms of $k$. Then, assuming a trivial beta-function
with one term (c=0), the renormalization group fixes the asymptotic
behaviour for general $\mu^{2}$,
\newpage
\be
d_{k}\approx A\left(\frac{\mu^{2}}{Q^{2}}\right)^{\h{bz_{i}}{2}}
\Biggr(\frac{1}{z_{i}}\Biggr)^{k}k!(\mbox{ })\;.
\ee
So such a singularity at $z=z_{i}$ corresponds to $Q^{-bz_{i}}$
scaling behaviour. To reproduce the $Q^{-4},Q^{-6},\ldots$ behaviour
of the OPE condensates one then requires Borel plane singularities
at $z=4/b,6/b,\ldots$, the IR renormalons.

In QED, one-loop vacuum polarization diagrams with a chain of vacuum
polarization bubbles inserted lead to fixed-sign factorial growth and
UV renormalon singularities \cite{q}. These are associated with the
Landau
pole in QED. For the QCD vacuum polarization function one can also
consider one-loop diagrams with a single gluon line inserted.
Applying a cut-off on the momentum of this line, inserting the
running QCD coupling and integrating over the high-momentum region,
Mueller \cite{s} has shown that one can explicitly derive the form
of the
leading UV renormalon at $z=-2/b$; and, by integrating over momenta
less than the cut-off, of the leading IR renormalon at $z=4/b$.

The final conclusion is that one expects UV renormalon
singularities, $\mbox{UV}_{\ell}$, at $z=z_{\ell}=-2\ell/b$, $\ell=1
,2,3,\ldots$; and IR renormalon singularities, $\mbox{IR}_{\ell}$,
at $z=z_{\ell}=2\ell/b$, $\ell=1,2,3,\ldots$. For the specific case
of QCD vacuum polarization one expects $\mbox{IR}_{1}$ to be absent
since, as discussed, there is no dimension two condensate.

These singularities at $z=z_{\ell}$ in the Borel plane should be
branch points of the form \cite{s}
\be
B[\tilde{D}](z)=\frac{A_{0}+A_{1}(1-z/z_{\ell})+
O((1-z/z_{\ell})^{2})}{(1-z/z_{\ell})^{p+cz_{\ell}}}\;,
\ee
where $p$ is a positive integer. The exponent $cz_{\ell}$ is fixed
by the renormalization group in order to give the required $Q^{2}$
scaling, taking into account a realistic two-term beta-function with
$c\neq 0$. $\mbox{UV}_{1}$ should be the singularity closest to the
origin (assuming that $\mbox{IR}_{1}$ is absent) and so will give
the overall asymptotic large-order behaviour of the $d_{k}$
coefficients. Corresponding to a branch point in the Borel plane of
the form of equation (27), one has the large-order behaviour
\be
d_{k}=\frac{A_{0}}{\Gamma(p+cz_{\ell})}\left(\frac{1}{z_{\ell}}
\right)^{k}\Gamma(k+p+cz_{\ell})\;.
\ee
So for $\mbox{UV}_{1}$ we have the leading behaviour
\be
d_{k}\approx\frac{A_{0}}{\Gamma(p-2c/b)}\left(-\frac{b}{2}\right)
^{k}k^{p-1}k^{-2c/b}k!(1+O(1/k))\;,
\ee
where we have expanded the second $\Gamma$-function of equation
(28) for large $k$. The $A_{1}$ and higher terms in the numerator
Taylor series about $z=z_{\ell}$ in equation (27) are O($1/k$)
sub-asymptotic effects. The explicit diagrammatic evaluations
\cite{f,g,s}
give $A_{0}=\frac{4}{9}\Gamma(2-2c/b)$, $p=2$ for $\mbox{UV}_{1}$,
in the MOM scheme with $\mu^{2}=Q^{2}$.

To make contact with the $1/N_{f}$ and $1/b$ expansions of equations
(9) and (12) we note that from (8)
\be
\frac{c}{b}=\left[-\frac{7}{8}\frac{C_{A}^{2}}{b^{2}}+\frac{5}{4}
\frac{C_{A}}{b}-\frac{11}{8}\frac{C_{A}C_{F}}{b^{2}}+\frac{3}{4}
\frac{C_{F}}{b}\right]\;.
\ee
Then writing $k^{-2c/b}$ as $e^{-2c\ln k/b}$ and expanding we have
\ba
d_{k}&\approx&\frac{A_{0}}{\Gamma(p-2c/b)}\left(-\frac{b}{2}\right)
^{k}k^{p-1}k!\biggr[1-\frac{5}{2}\frac{C_{A}}{b}\ln k-\frac{3}{2}
\frac{C_{F}}{b}\ln k\nonumber\\& &+\left(\frac{25}{8}\ln^{2}k
+\frac{7}{4}\ln k\right)\frac{C_{A}^{2}}{b^{2}}+\ldots\biggr]\;.
\ea
We shall define
\be
\frac{A_{0}}{\Gamma(p-2c/b)}=A_{01}+\frac{A_{02}}{b}+\frac{A_{03}}
{b^{2}}+\ldots\;,
\ee
where $A_{0r}$ is a sum of multinomials in $C_{A}$ and $C_{F}$ of
degree
$r-1$ and $A_{01}$ is a pure number. One can then
obtain the asymptotic behaviour of the coefficients to all
orders in the $1/N_{f}$ and $1/b$ expansions. The $1/b$ expansion
corresponds to the successive terms in equation (31). So we have
\ba
d_{k}^{(k)}&\approx&A_{01}\left(-\frac{1}{2}\right)^{k}k^{p-1}k!
\nonumber\\
d_{k}^{(k-1)}&\approx&A_{01}\left(-\frac{1}{2}\right)^{k}k^{p-1}k!
\left(-\frac{5}{2}C_{A}\ln k-\frac{3}{2}C_{F}\ln k\right)\nonumber\\
\vdots& &
\ea
For the $1/N_{f}$ expansion asymptotics the leading terms will come
from the binomial expansion of $b^{k}=
(\frac{11C_{A}-2N_{f}}{6})^{k}$,
\be
d_{k}^{[k-r]}\approx A_{01}\frac{6^{-k}}{r!}\left(-\frac{11}{2}
C_{A}\right)^{r}k^{p+r-1}k!
\ee
where $r=0,1,2,\ldots$. Note that an
additional factor of $k$ is gained for each additional order in the
$1/N_{f}$ expansion.
The leading term in the large-$N$ expansion, $d_{k}^{[0]}$, is
given by
\be
d_{k}^{[0]}\approx\tilde{A}_{01}12^{-k}(-11C_{A})^{k}k^{p-1}
k^{-102/121}k!
\ee
where $\tilde{A}_{01}$ is the large-$N$ limit of equation (32). Of
course $\tilde{A}_{01}$ cannot be determined from the large-$N_{f}$
result.

Assuming a $\mbox{UV}_{1}$ singularity as in equation (27) we
therefore apparently have the remarkable conclusion that, given an
exact large-$N_{f}$ (large-$b$) result for $d_{k}^{[k]}$
($d_{k}^{(k)}$), we can obtain $A_{01}$ and $p$; and hence the
asymptotics (up to an O($1/k$) correction) of the coefficients to
{\em all orders} in the $1/N_{f}$ ($1/b$) expansion are determined
by equations (33) and (34). We shall check in due course that the
exact large-$N_{f}$ result of equation (18) gives $A_{01}=4/9$ and
$p=2$ (MOM scheme, $\mu^{2}=Q^{2}$) in agreement with other
evaluations \cite{f,g,s}.

The conclusion that the large-$N_{f}$ result can determine the full
asymptotics beyond leading order in $1/N_{f}$ seems too good to be
true; and indeed recent work by Vainshtein and Zakharov \cite{g}
casts
doubt on it. These authors have systematically developed a ``UV
renormalon calculus'' in which the leading ultra-violet behaviour of
loop diagrams with different numbers of chains of vacuum polarization
graphs inserted is extracted using an operator product expansion.
Including one chain they reproduce a $\mbox{UV}_{1}$ result of the
form of equation (27) in agreement with the result quoted above.
Explicitly evaluating the two-chain (three-loop) result in a
simplified U(1) model they find a contribution to the UV renormalon
asymptotics which is $1/N_{f}$ down on the one-chain result; but has
additional powers of $k$ and so dominates the one-chain result.
Conventional wisdom would have expected additional chains to have a
suppression by powers of $k$ and hence not to contribute to the
$A_{0}$ coefficient in equation (27). Combinatoric factors conspire
to modify this, however.

The UV renormalon calculus results \cite{g} imply that the Borel
plane
singularity at $z=-2/b$ in equation (27) should be modified to
\be
B[\tilde{D}](z)=\sum_{m=1}^{\infty}\frac{A_{0}^{(m)}+A_{1}^{(m)}
(1+bz/2)
+O((1+bz/2)^{2})}{(1+bz/2)^{\gamma_{m}-2c/b}}
\ee
where $m$ labels the number of vacuum polarization chains, $m=1$
corresponding to the previous one-chain results. Counting powers of
$N_{f}$, one expects each additional chain to give a $1/N_{f}$
suppression and hence equation (32) should generalise to
\be
\frac{A_{0}^{(m)}}{\Gamma(\gamma_{m}-2c/b)}=\frac{A_{01}
^{(m)}}{b^{m-1}}
+\frac{A_{02}^{(m)}}{b^{m}}+\ldots
\ee
The exponent $\gamma_{m}$ will be related to the anomalous dimensions
of the operators appearing in the operator product expansion of
reference \cite{g}. There will be a number of operators and hence a
number of contributions to equation (36) for any $m$. Each such
$\gamma_{m}$ will have a $1/N_{f}$, $1/b$ expansion and we can define
the leading term $p_{m}$, typically a positive integer. We select for
each $m$ the $\gamma_{m}$ contribution with largest $p_{m}$; and this
is the single term displayed in equation (36).
Providing that $p_{m+1}>p_{m}+1$ for any $m$,
then (36) and (37) lead to the $1/N_{f}$, $1/b$ expansion
asymptotics
\ba
d_{k}^{[k-r]}&\approx&A_{01}^{(r+1)}\frac{(-3)^{r}}{6^{k}}
k^{p_{r+1}-1}
k!\nonumber\\
d_{k}^{(k-r)}&\approx&A_{01}^{(r+1)}\left(-\frac{1}{2}\right)^{k}
k^{p_{r+1}-1}k!
\ea
with $r=0,1,2,\ldots$. The $A_{01}^{(r+1)}$ will consist of
multinomials in $C_{A}$, $C_{F}$ of degree $r$.
The inequalities on $p_{m}$ are required
since,
from equation (34), we see that, at fixed $m$, one gains an extra
factor of $k$ for each additional $1/N_{f}$ order.
The results of (38) imply that one does not get something for nothing
after all; but requires O($(1/N_{f})^{r}$) results (diagrams with
$r+1$ chains) to obtain the leading asymptotics to this order in the
$1/N_{f}$ expansion.

Having discussed the general expectations for the Borel plane
singularity structure we return to the exact large-$N_{f}$ result of
equation (18) and exhibit what its singularity structure actually is.

\section{Borel Plane Singularities of the Exact Large-$N_{f}$ Result}

Returning to equation (15) we can use Taylor's theorem to write
\be
\Psi_{n}^{[n]}=\frac{3^{2-n}}{2}(n-2)!\,\BC_{n-2}\tilde{P}(u)
\ee
where $\BC_{n}F(x)$ denotes the coefficient of $x^{n}$ in the
expansion
of $F(x)$ as a power series.
\be
\tilde{P}(u)=\frac{32}{3(2+u)}\sum_{k=2}^{\infty}\frac{(-1)^{k}k}
{(k^{2}-(1+u)^{2})^{2}}
\ee
which is obtained from equation (16) by writing $u=x-1$. Using
partial fractions on equation (40) and isolating the required
coefficient one finds the rather simple result
\ba
\frac{\Psi_{n}^{[n]}}{(n-2)!} &=& \frac{6n-1}{3^{n+1}}
-\frac{6n+1}{6^{n+1}}+12\sum_{\ell=3}^{\infty}
(-1)^{\ell+1}\biggr\{n\biggr(\frac{1}{\ell+1}
-\frac{1}{\ell+2}\biggr)\nonumber\\
& &+\biggr(-\frac{1}{(\ell+1)^{2}}+\frac{2}{(\ell+2)^{2}}
\biggr)\biggr\}\biggr(\frac{1}{3\ell}\biggr)^{n}\nonumber\\
& &-\frac{3}{(-6)^{n-1}}+12\sum_{\ell=3}^{\infty}(-1)^{\ell+1}
\biggr\{n
\biggr(\frac{1}{\ell-1}-\frac{1}{\ell-2}\biggr)\nonumber\\
& &+\biggr(\frac{1}{(\ell-1)^{2}}
-\frac{2}{(\ell-2)^{2}}\biggr)\biggr\}\biggr(\frac{-1}{3\ell}
\biggr)^{n}\;.
\ea
Putting in factors of $T_{f}^{n}N_{f}^{n}=(1/2)^{n}N_{f}^{n}$ and
replacing $N_{f}$ by $(-3b)$, we see that successive terms in
equation (41) are proportional to $(-b/2)^{n}$, $(-b/4)^{n}$,
$(-b/2\ell)^{n}$, $(b/4)^{n}$ and $(b/2\ell)^{n}$ corresponding
exactly to
the poles $\mbox{UV}_{1}$, $\mbox{UV}_{2}$, $\mbox{UV}_{\ell}$,
$\mbox{IR}_{2}$ and $\mbox{IR}_{\ell}$ respectively in the Borel
plane. There is no term involving $(1/(-3))^{n}$ and so
$\mbox{IR}_{1}$ is indeed absent as anticipated. The linear factor
proportional to $n$ in all but $\mbox{IR}_{2}$ means that all the
poles are double poles ($p=2$) except for $\mbox{IR}_{2}$ which is a
simple pole ($p=1$). Notice that the branch point exponent
$cz_{\ell}$ is sub-leading in the $1/N_{f}$, $1/b$ expansion and so
one sees poles and not branch points in the large-$N_{f}$ limit.

\newpage
Using equation (18) we can then obtain the leading-$b$ result for
$d_{k}$ ($\overline{\mbox{MS}}$ scheme, $\mu^{2}=Q^{2}$):
\ba
d_{k}^{(k)}b^{k}&=&k!\biggr\{\sum_{\ell=1}^{\infty}[A_{0}(\ell)
k+A_{1}(\ell)]
\biggr(-\frac{b}{2\ell}\biggr)^{k}\nonumber\\
& &+\sum_{\ell=1}^{\infty}[B_{0}(\ell)k+B_{1}(\ell)]\biggr(\frac{b}
{2\ell}
\biggr)
^{k}\biggr\}
\ea
where
\ba
A_{0}(1)&=&\frac{4}{9}e^{-5/3}\mbox{;      }A_{1}(1)=\frac{14}{9}
e^{-5/3};
\nonumber\\
A_{0}(2)&=&-\frac{1}{3}e^{-10/3}\mbox{;      }A_{1}(2)=-\frac{11}
{6}e^{-10/3};\nonumber\\
A_{0}(\ell)(\ell\geq 3)&=&\frac{8}{3}\frac{1}{\ell^{2}}
\left(\frac{1}{\ell+1}+
\frac{1}{\ell+2}\right)e^{-5\ell/3};\nonumber\\
A_{1}(\ell)(\ell\geq 3)&=&\frac{8}{3}\frac{1}{\ell^{2}}
\biggr[\left(2+\frac
{5\ell}{3}\right)\left(\frac{1}{\ell+1}+\frac{1}{\ell+2}\right)
\nonumber\\
& &+\left(
-\frac{1}{(\ell+1)^{2}}+\frac{2}{(\ell+2)^{2}}\right)\biggr]e^
{-5\ell/3};
\nonumber\\
B_{0}(1)&=&B_{1}(1)=0;\nonumber\\
B_{0}(2)&=&0\mbox{;      }B_{1}(2)=e^{10/3}\nonumber\\
B_{0}(\ell)&=&-A_{0}(-\ell)\mbox{;      }B_{1}(\ell)=-A_{1}(-\ell)
\mbox{  }
(\ell\geq 3)\;.
\ea
To obtain the result in the MOM scheme with $\mu^{2}=Q^{2}$ one
simply omits the exponential factors in equations (43). The first sum
in equation (42) generates $\mbox{UV}_{\ell}$ singularities of the
form of equation (27) and the second sum the $\mbox{IR}_{\ell}$
singularities. $\mbox{IR}_{1}$ is absent ($B_{0}(1)=B_{1}(1)=0$)
as required from the OPE;
and all poles are double ($p=2$) except $\mbox{IR}_{2}$ for which
$B_{0}(2)=0$ giving a simple pole ($p=1$).
For $\ell\geq 3$ there is a curious and unexplained
symmetry between the residues of $\mbox{UV}_{\ell}$ and
$\mbox{IR}_{\ell}$ with $A(\ell)=-B(-\ell)$. As promised the
coefficient
$A_{0}(1)=\frac{4}{9}e^{-5/3}$ is in agreement with other
calculations of $\mbox{UV}_{1}$ \cite{f,g,s}. The coefficient
$B_{1}(2)=e^{10/3}$ is consistent with the result of \cite{s}
for the first
IR renormalon, $\mbox{IR}_{2}$.

Since equation (41) is split into contributions from UV and IR
renormalons we can sum up these contributions separately. Summing
the first three UV terms in (41) one obtains
\ba
\frac{\Psi_{n}^{[n]}}{(n-2)!}(\mbox{UV}) &=& -\frac
{2(\zeta_{2}-2)}{(-3)^{n-1}}
-\frac{2\zeta_{2}-3}{(-6)^{n-1}}\nonumber\\
& &+\frac{4}{(-3)^{n-1}}\sum_{m=1}^{n-1}(-1)^{m}m
(1-2^{-m})(1-2^{m-n})\zeta_{m+1},
\ea
and summing the final two IR terms gives
\ba
\frac{\Psi_{n}^{[n]}}{(n-2)!}(\mbox{IR}) &=& -\frac{2(n^{2}
-3n+4-\zeta_{2})}
{(-3)^{n-1}}-\frac{\frac{1}{2}(n^{2}+3n+2-4\zeta_{2})}{(-6)
^{n-1}}\nonumber\\
& &+\frac{4}{(-3)^{n-1}}\sum_{m=1}^{n-1}m
(1-2^{-m})(1-2^{m-n})\zeta_{m+1}.
\ea
The UV and IR pieces separately contain even and odd
$\zeta$-functions but the even $\zeta$-functions cancel in the sum of
(44) and (45) to reproduce equation (17) which contains only odd
$\zeta$-functions.

We finally discuss the connection between the Borel plane singularity
structure of $\tilde{D}$, which we have discussed extensively, and
that of the more experimentally-relevant $\tilde{R}$, related to it
by the dispersion relation of equation (4). It is easy to show that
in the large-$b$ limit ($c=0$) \cite{k}
\be
B[\tilde{R}](z)=\frac{\sin(\pi bz/2)}{\pi bz/2}B[\tilde{D}](z)
\ee
Since $\sin(\pi bz/2)$ has single zeros $\sim(z-z_{\ell})$ at the
same positions as the renormalon singularities one finds that
renormalon poles of order $p$ in $B[\tilde{D}]$ are converted to
poles of order $p-1$ in $B[\tilde{R}]$. This implies that the poles
in $B[\tilde{R}]$ are simple poles except for $\mbox{IR}_{2}$ which
was a simple pole in $B[\tilde{D}]$ and hence apparently vanishes
\cite{f}.
The absence of the $\mbox{IR}_{1}$ singularity at $z=2/b$ in
$B[\tilde{D}](z)$ implies from equation (46) that $B[\tilde{R}](z)$
must have a compensating zero at this position. Brown and Yaffe
\cite{k}
considered this unlikely and hence cast doubt on the absence of
$\mbox{IR}_{1}$. The exact large-$N_{f}$ result shows that there is
indeed no $\mbox{IR}_{1}$ singularity in $B[\tilde{D}]$ and hence
such a zero is present in $B[\tilde{R}]$.

The leading asymptotics of the coefficients $r_{k}$ will be given by
$\mbox{UV}_{1}$ for $B[\tilde{R}](z)$. Expanding around $z=-2/b$ we
have
\be
\frac{\sin(\pi bz/2)}{\pi bz/2}=\biggr(1+\frac{bz}{2}\biggr)+O\Biggr(
\biggr(1+\frac{bz}{2}\biggr)^{2}\Biggr)
\ee
and hence from equation (46) we find that the asymptotic behaviour of
$r_{k}$ is given by changing $p\rightarrow p-1$ in the results for
$d_{k}$ (equations (33) and (34)). Even assuming the more complicated
$\mbox{UV}_{1}$ structure of equation (36) one simply changes $p_{m}
\rightarrow p_{m}-1$. This implies that on very general grounds one
expects
\be
\frac{r_{k}}{d_{k}}\approx\frac{1}{k}(1+O(1/k))
\ee
so that the $r_{k}$ coefficients grow more slowly asymptotically.

We conclude by noting that exact large-$N_{f}$ results do exist also
for other QCD observables. In particular, reference \cite{i}
contains a
leading-$N_{f}$ result for the perturbative coefficients of the
radiative corrections to the
Gross--Llewellyn-Smith (GLS) sum rule. This reveals that in the Borel
plane the GLS corrections have simple poles at $z=\pm 2/b,\pm 4/b$.
So the first two UV and IR renormalons are present but the remaining
renormalons are absent to leading order in $1/N_{f}$. It would be
interesting to compare this with the OPE structure for these
corrections.

\section{Vanishing IR Renormalons}

The singularity structure in the Borel plane for SU($N$) QCD will
change as $N$ and $N_{f}$ are varied. In particular, if $N_{f}$
approaches $11N/2$ from below then $b=(11N-2N_{f})/6$ will
approach zero from above; and, as more flavours of quark are added,
the $\mbox{IR}_{\ell}$ and $\mbox{UV}_{\ell}$ singularities, which
are spaced at intervals of $2/b$, will move outwards away from the
origin in the $z$-plane. The instanton/anti-instanton
($\mbox{I}\,\overline{\mbox{I}}$) singularities remain fixed at
$z=4,8,12,\ldots$
independent of $N$ and $N_{f}$. When $b=1/2$ ($N_{f}=15$ for SU(3)),
$\mbox{IR}_{1}$ will be at the same position as the leading
$\mbox{I}\,\overline{\mbox{I}}$ singularity at $z=4$; and for
flavour
saturation (maximum $N_{f}$ for which $b>0$, $N_{f}=16$ for SU(3))
$b=1/6$ and the leading singularity on the positive axis will be the
$\mbox{I}\,\overline{\mbox{I}}$ at $z=4$.

For SU(3) QCD a remarkable phenomenon first occurs at $b=1/2$
($N_{f}=15$). One finds that the branch point exponent of
$\mbox{IR}_{\ell}$, $2c\ell/b$ in equation (27), becomes a negative
integer and the structure of $\mbox{IR}_{\ell}$ in the Borel plane
is then
\be
B[\tilde{D}](z)=A_{0}\biggr(1-\frac{z}{4\ell}\biggr)^{88\ell-p}\;.
\ee
The $\mbox{IR}_{\ell}$ singularity disappears provided that
$p<88\ell$.
For the particular case of the D-function we have apparently $p=1$ or
2 and so all of the IR renormalons disappear at $b=1/2$ and the only
singularities on the positive $z$-axis are those due to instantons.
Notice that the UV renormalons are still present but become poles.

As
first noted by White \cite{l,m} the IR renormalon singularities also
disappear for $N_{f}=16$ flavour-saturated SU(3). The exponent
$2c\ell/b$ again becomes a negative integer and the
$\mbox{IR}_{\ell}$ structure in the Borel plane is
\be
B[\tilde{D}](z)=A_{0}\biggr(1-\frac{z}{12\ell}\biggr)^{906\ell-p}\;.
\ee
So for $p<906\ell$ the $\mbox{IR}_{\ell}$ singularities disappear.

The implication is that $N_{f}=15$ and 16 SU(3) QCD are very
special instanton-dominated theories. At precisely the point when
instantons become the leading singularities all the other IR
renormalon singularities on the positive $z$-axis vanish. It is
interesting to ask if this scenario is unique to SU(3) or can be
realised for other values of $N$. $b=1/2$ will occur for integer
$N_{f}$ only for $N$ odd. For $N$ even the
$\mbox{I}\,\overline{\mbox{I}}$ singularity becoming leading and
flavour saturation are telescoped into the single value $b=1/3$.

Considering $N$ odd first, we require that $c/b$ is integer for
$b=1/2$;
\be
\biggr(\frac{c}{b}\biggr)\biggr|_{b=\h{1}{2}}
=\frac{(-25N^{3}+13N^{2}+11N-3)}{4N}\;.
\ee
A necessary condition for this to be an integer is that $3\equiv 0$
$(\mbox{mod } N)$, which uniquely fixes $N=3$. Then $(c/b)|_{b=
\frac{1}
{2}}=-44$.

For $N$ odd and flavour saturation $b=1/6$;
\be
\biggr(\frac{c}{b}\biggr)\biggr|_{b=\h{1}{6}}
=\frac{(-225N^{3}+39N^{2}+99N-9)}{4N}\;.
\ee
A necessary condition for this to be an integer is that $9\equiv 0$
$(\mbox{mod } N)$ so $N=3$ or 9. The $N=9$ case gives
$(c/b)|_{b=\frac{1}{6}}=-4444$; and, for $N=3$,
$(c/b)|_{b=\frac{1}{6}}=-453$.

For $N$ even and $b=1/3$
\be
\biggr(\frac{c}{b}\biggr)\biggr|_{b=\h{1}{3}}
=\frac{(-225N^{3}+78N^{2}+99N-18)}{4N}\;.
\ee
A necessary condition for this to be an integer is that $18\equiv 0$
$(\mbox{mod } N)$ and so $N=2,6$ or 18. $N=6$ is the only case for
which $c/b$ is an integer and then $(c/b)|_{b=\frac{1}{3}}=-1884$.

So for flavour-saturated SU($N$) IR renormalons are absent only for
$N=3,6$ or 9. For SU(9) the $b=1/2$ case where the
$\mbox{I}\,\overline{\mbox{I}}$ singularity becomes leading still has
IR renormalons.

We conclude that SU(3) QCD is a very special theory from yet another
point of view.
\section{Conclusions}
In this paper we have discussed and sought to extend our knowledge of
the Borel plane singularity structure of the Adler D-function (QCD
vacuum polarization). This singularity structure succinctly encodes
the large-order asymptotic behaviour of the perturbation theory
coefficients.

We pointed out that an expansion of the perturbative coefficients in
powers of $b$, the first QCD beta-function coefficient, rather than
in $N_{f}$, was natural when comparing with QCD renormalon
expectations. We further noted that, if the leading UV renormalon
indeed has the expected structure of a simple branch point, then
knowledge of the perturbative coefficients to leading order in
$1/N_{f}$, $1/b$ allows the large-order behaviour to all-orders in
$1/N_{f}$, $1/b$ to be inferred, equations (33) and (34). This
seems unlikely and a more complicated structure was proposed,
equation (36), which is consistent with the UV renormalon calculus
results of Vainshtein and Zakharov \cite{g}.

Using the exact large-$N_{f}$ result for the D-function of reference
\cite{i} we exhibited the explicit
singularity structure in the $z$-plane and found the expected UV
and IR renormalon singularities. They appear as poles in the
large-$N_{f}$ limit. In particular, the first IR renormalon, which
would correspond to $Q^{-2}$ behaviour not present in the OPE, is
absent. We gave explicit expressions for the residues at all of
these poles (equations (43)) and unexpectedly found those for the
third and higher UV and IR renormalons to be symmetrically related.
We were also able to sum up separately the UV and IR renormalon
contributions in closed form (equations (44) and (45)) and obtained
expressions
containing even and odd $\zeta$-functions. When UV and IR
contributions are combined the even $\zeta$-functions cancel.

We finally noted that in flavour-saturated SU($N$) QCD the IR
renormalons are absent for $N=3,6$ and 9. These theories then have
ambiguities dominated by instantons. For SU(3) the IR renormalons
first disappear when $N_{f}=15$ ($b=1/2$), at which point the
$\mbox{I}\,\overline{\mbox{I}}$ singularity is in the same
position as $\mbox{IR}_{1}$ and becomes leading.

\section*{Acknowledgements}

We would like to thank Alan White for stimulating and interesting
discussions about the disappearance of IR renormalons. C.L-T
gratefully acknowledges the receipt of a P.P.A.R.C. U.K. Studentship.

\end{document}